# BIMS: Biomedical Information Management System

## Oscar Mora[a], Jesús Bisbal[a]

[a]*Universitat Pompeu Fabra, Barcelona, Spain*

## Abstract

*In this paper, we present BIMS (Biomedical Information Management System). BIMS is a software architecture designed to provide a flexible computational framework to manage the information needs of a wide range of biomedical research projects. The main goal is to facilitate the clinicians' job in data entry, and researcher's tasks in data management, in high data quality biomedical research projects. The BIMS architecture has been designed following the two-level modeling paradigm, a promising methodology to model rich and dynamic information environments.*

*In addition, a functional implementation of BIMS architecture has been developed as a web-based application. The result is a highly flexible web application which allows modeling and managing large amounts of heterogeneous biomedical data sets, both textual as well as visual (medical images) information.*

## Introduction

The scientific research in the biomedical field has experienced a spectacular growth last years [4]. Advances in analytical technology have produced an enormous quantity of biomedical information [1]. Any biomedical researcher has the possibility to access to lots of real data to test his investigations, and obtain more reliable results. Unfortunately, the challenges to appropriately manage this biomedical information have also experienced a spectacular leap [5]. Usually, tasks like acquire, store, or analyze data are difficult due to their heterogeneity [1]. Moreover, there is not a flexible computational solution to address completely the needs which commonly arise in a biomedical research project, based on gathering patient information. In contrast to clinical trials, where data quality is a critical aspect of the project, the main goal here is to facilitate all tasks related to biomedical data capturing and their management, providing good levels information quality [5]. The ultimate goal should be to develop a flexible framework to reduce the time spent in data entry, facilitating the job of the clinician, and the analytical tasks of the researchers.

This paper presents a software architecture called BIMS (Biomedical Information Management System) and its functional implementation in a web-based software application.

## Motivation

From the information management point of view, biomedical research projects consist of capturing, transporting, and managing (in a general sense) large amounts of heterogeneous information, related to a specific clinical specialty. Figure 1 shows the three-phase information workflow. In the context of this paper, this is always understood as the management of patient information. This workflow can be understood as a pipeline, where the information enters the workflow at the initial (capture) phase, and travels to the end (managing) phase. In the last phase, the biomedical information is available to all related people who take part in the project.

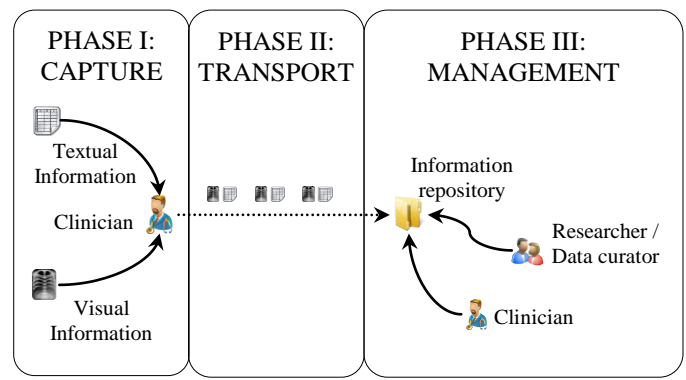

*Figure 1: The three-phase information workflow*

The following sections describe these phases in details.

### Phase I: Capture

Capture is the first phase of the information workflow. In this step, information about a patient, relevant to a biomedical research study, is gathered using a well-defined protocol. Typically, biomedical information gathered is divided in two groups, on the one hand, visual information as images (e.g. x-ray, ultrasound, computerized topography). On the other hand, textual information like patient measurements (e.g. volume of left ventricle), or demographic information.

The capture phase is the main entry of the information to the biomedical research study, due to this, it is necessary to use a validation mechanism to reject invalid data.

**Phase II: Transport**

Transport is the second phase of the workflow. In this step all biomedical information captured in the first phase, is moved to a structured information repository that will subsequently allow for this information to be considered in one or more biomedical studies. Due to privacy issues it is necessary to define a secure mechanism to protect the critical information from the external threats.

**Phase III: Management**

The last phase of the workflow is to manage the biomedical information that has been captured and transported in the previous phases. Managing is a global term to describe a set of functionalities to check, store, and access to the biomedical information, in different ways, depending on the role of the person who accesses, and the objectives of each biomedical study.

**Current practice**

Currently, the methodologies used to implement the three-phase information workflow are largely based on manual procedures [5] like recording textual information in paper forms, burning images in DVD's, and sending them physically to the research facility. Other possibilities are based on immature and ad-hoc software applications which do not provide a complete infrastructure to capture, transport, and manage textual and visual information in a sufficiently flexible manner [5].

In summary, the main motivation is to provide a computational system to manage the three-phase workflow information of a biomedical research project. Thus, the goals are: (1) to define and model biomedical concepts, and group them in sections according to their meaning; (2) to generate, on the fly, user friendly forms in order to request patient information of biomedical concepts defined previously; (3) to transport the information captured in the forms, to the information repository, in a secure way; (4) to store, manage, and display the information captured in several ways, depending on the role of the user (clinicians, researchers, or data curators).

## System architecture

The solution developed to fulfill the workflow described above is BIMS (Biomedical Information Management System). BIMS is a modular multilayer object–oriented software architecture, designed with to provide a flexible infrastructure to capture, and manage large amounts of heterogeneous biomedical data sets, both textual as well as visual information.

Traditionally, the software engineering approach to modeling a domain is based on creating a software model to map the domain elements in, an almost, *one-to-one* relationship [2]. The resulting software model is composed of a large set of objects, which contains all the information that is considered necessary for the software application. The necessary information is identified, and gathered by the software architects, and the user-domain experts who take part in the project development [3]. Hence, business entities are modeled directly in software and database models, via an iterative process of writing use cases, finding software objects, and building models which will eventually become software [10]. Moreover, in relational database developments, concepts are encoded in the relational database schema and into program code. As a result, in a majority of object-oriented and relational systems, the semantic concepts are nearly hard-coded [2]. This approach, typically named *single-level modeling*, has repeatedly been followed in the biomedical domain, and it has proved rather unsatisfactory [7]. The reason is that the biomedical domain is characterized [3, 9] by being:

- *Large*. A well known ontology: SNOMED-CT (www.ihtsdo.org) contains over 350.000 atomic concepts and 1.5 million relationships.

- *Complex*. Different views of information, requirements, and granularity.

- *Open − ended*. The numerous advances in biomedical research constantly update clinical practice. This, in turn, requires the underlying information system to be constantly maintained.

With the aim to design a very flexible system, BIMS architecture is based on a evolution of the single-level modeling, which led to the so-called *two-level modeling* paradigm [2, 8].

**Two level modeling paradigm**

The goal of this alternative modeling methodology is to separate the semantics of the *information* and the *knowledge* into two levels of the model [2]. The first level is composed of a set of simple and abstract meaning software objects. It is named *Reference Model*. This first level must be small in size, to be comprehensible, and contain only non-volatile concepts to be maintainable. The second level encapsulates the knowledge of the domain in specific volatile concepts, defined using a set of flexible rules. Thus, *two-level modeling* is based on a set of building blocks in the first level, organized in a hierarchical structure according to a set of generic constraints [3], established at run-time. The flexibility of the first level software objects, and the constraints defined in the second level allows modeling a numerous set of structures and volatile concepts of *large, complex* and *open-ended* domains, without having to update the underlying information infrastructure.

**Two level modeling in BIMS architecture**

In order to provide the possibility of managing a wide range of biomedical research projects, the BIMS architecture has been designed using the two-level modeling paradigm. On the one hand, the first level of the BIMS design is composed of a reduced set of abstract meaning software objects (*BIMS Reference Model*), which provide a wide range of combination possibilities to represent non-volatile concepts. The second level is a set of flexible rules to define and constrain the combinations of first level software objects. Thus it is possible to ex-press an extensive set of non-volatile concepts of the modeled domain.

### First level: BIMS Reference Model

Following the single level modeling paradigm to represent the domain elements, it would be necessary to set, describe, and study every parameter requested in the biomedical research project, as well as their internal organization in sections, or subsections. For instance, if the biomedical research project needs to manage the *blood pressure*, it would be necessary to define a software object to represent the domain concept of the blood pressure, with its proper restrictions (e.g. avoid automatically a pressure of *1755* mm Hg). Such an approach is not sustainable in a dynamic environment as is the case of the biomedical research information domain.

Using the two-level modeling paradigm, in contrast, it is necessary to detect, and model the non-volatile (constant in time) concepts of the domain. In this case, the biomedical research project can be managed as a set of requested medical parameters, organized in sections. A medical parameter is an element which contains a set of values. It is important to note that only a small number of facts are concretely modeled in this way. Figure 2 shows an UML diagram which represents the most important objects of BIMS Reference Model:

- *MedicalParameter*. This object can represent any kind of biomedical concept requested in the biomedical study.

- *Value*. This object can represent an alphanumerical string, or DICOM image reference to a PACS server, associated to a *MedicalParameter*.

- *Section*. This object is used as a hierarchical organizing method. It can contain other objects of the same type, or *MedicalParameter* elements.

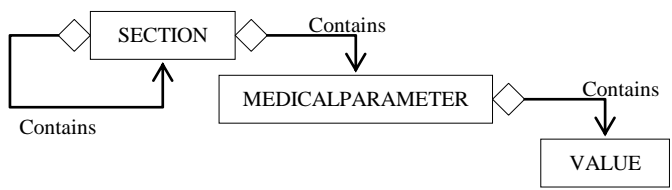

*Figure 2: BIMS Reference Model diagram*

As stated in *Motivation* section, it is necessary to use a mechanism to ensure an acceptable level of data quality. *MedicalParameter* allows controlling the content of the *Value* object, by specifying a maximum / minimum numerical threshold or forcing to the content to match with a specific regular expression.

In the case of visual information, the *MedicalParameter* object provides the possibility to define a PACS server connection to send DICOM file. The *Value* object stores the DICOM file reference to request the same image to PACS server.

In summary, the BIMS Reference Model is composed of a set of simple software objects which represent concepts without semantic load. These software objects can be to build hierarchical structures to represent volatile concepts of the domain.

### Second level: rules to define the knowledge

The software objects of the BIMS Reference Model can be organized in complex hierarchical structures with semantic load. In this way, the mission of the second level is to provide a flexible framework, based on a set of rules, to specify the available combinations of the software objects, as well as their properties. The combination of the software objects allows representing volatile concepts of the biomedical domain.

The BIMS architecture provides a mechanism to define the available combinations of the first level software objects, by using a set of text rules in a well defined schema. The rules allow defining properties and relationships of the software objects of BIMS Reference Model..

Hence, these rules allow modeling the knowledge of a biomedical study, at run-time. For instance, it provides the possibility to use a *MedicalParameter* object to represent the height of a patient, and define its associated *Value* object as a numerical element in [80, 260] range. It is interesting to note the clear separation between information and knowledge, produced by two level modeling. On the one hand, BIMS Reference Model only provides a couple of software objects to represent a numerical value, without semantic load. On the other hand, the second level provides a textual rule to use the *MedicalParameter* software object to represent a biomedical concept like the patient height, with the proper restrictions.

In summary, the second level of BIMS architecture is a schema which allows using a set of textual rules to organize, at run-time, the software objects of the BIMS Reference Model, their relationships and their properties.

### Patient follow-up and anonymity mechanisms

In order to provide patient-level granularity, that is, able to identify all biomedical information of a patient who has taken part in several projects, it is necessary to manage every patient as a unique entity. This feature is useful to study time related aspects of a patient, like the evolution of a disease.

The goal is to generate a unique identifier to assign to the person who takes part in a biomedical project. In addition, it is necessary to detect if the same person takes part in various projects, to assign him the same identifier. Due to privacy reasons, it is not possible to use identifiable information (like name) to build the identifier. This information is named PHI (Personal Health Information), defined as "individually identifiable health information that is transmitted or maintained in any form or medium by a covered entity" [6] by the HIPAA (Health Insurance Portability and Accountability Act).

The BIMS architecture provides a patient identifier generation methodology based on a one-way function. Firstly, the clinician requests the national ID (D.N.I in Spain) to the person who is going to take part in the biomedical research project. Then, this national ID is used as a seed of a hash function, and the result is assigned to the person as his patient identifier. This method has two advantages. Firstly, it is not possible to find out the national ID from the generated patient identifier because the hash is a one-way mathematical function. Secondly, if the same person takes part in another project, the hash operation generates the same patient identifier, since the seed (national ID) is the same. In addition, the clinician has the possibility to elaborate a look-up table to recover the relationship between national ID and patient identifier of a person.

Visual information in DICOM files can contain PHI in the header fields. In this case, a similar methodology is applied to remove PHI, and provide the patient follow-up feature. The mechanism is based on seeking DICOM headers field which contains PHI, and overwrite them with the result of a hash operation. The seed of this hash operation is the previous content the header field. Thus, if two DICOM files have the same content in a header field, the result of hash it will be the same.

## Implementation

The BIMS architecture has been fully implemented in a web-based software application (www.tecn.upf.edu/~jbisbal/bims). Figure 3 shows the information workflow in the BIMS architecture implementation. The clinician uses a Web browser to enter biomedical information (textual information and DICOM images), which travels over a HTTPS (Hypertext Transfer Protocol Secure) connection to get the web-based application. There, textual information is stored in a DBMS (Database Management System), and DICOM images are sent to an external PACS server. The workflow design provides a specific access to the data curator, the researcher, and the clinician.

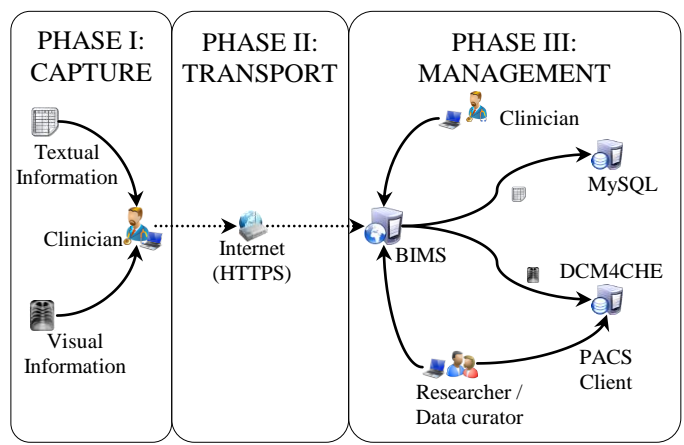

*Figure 3: Workflow of the BIMS architecture implementation*

The definition of the biomedical concepts is performed using a set of textual rules. To do this, an XML schema has been defined as framework to define such rules. The following XML code snippet shows an example definition of the *heart rate* as a biomedical concept, its properties (name, description), and its constraints (interval from 35 to 200]).

```
<name>Heart rate</name>

<min_value>35</min_value>

<max_value>200</max_value>
```

```
<used_unit>beat/m</used_unit>

<ontology>SNOMED-CT</ontology>

<ontology-code>364075005</ontology-code>
```

### Technologies

Firstly, two repositories have been used: MySQL (www.mysql.com) to store textual information, and DCM4CEE (www.dcm4che.org) as a PACS server to save DICOM files. Secondly, the Hibernate package (www.hibernate.org) is used as ORM (Object relational mapping) library. In addition, Spring framework (www.springsource.org) is used as infrastructure to implement the software application. At presentation level, JSF (Java Server Faces) framework (java.sun.com/javaee/javaserverfaces) and a set of Java Server Pages (JSP) are used to interact with the user. It is important to note that the BIMS architecture implementation is built on top of open source technologies.

## State of the art

One of the most interesting solutions which partially address the problems described above is *OpenClinica*.

OpenClinica (www.openclinica.org) is a free, open source clinical trial software platform for Electronic Data Capture (EDC) and medical data management in clinical research.

OpenClinica data model is based on a set of CDISC (www.cdisc.com) standard specifications, widely used for clinical trials management, but which lacks any kind of semantic (ontological) foundation. This limits its ability to provide advanced data processing, validations, and data quality control [5]. Furthermore, the current version of OpenClinica does not consider the collection of images in any way, which is an essential task in a biomedical research project. Moreover, OpenClinica does not support automatic mechanisms for managing patient follow-ups, to identify if the same patient has taken part in various biomedical research projects.

## Future work

The results of this project provide a strong foundation for the implementation of further ideas related to BIMS architecture.

### BIMS Query & Export Module

The information managed by BIMS is used by biomedical researchers in their investigations. Thus, a typical task of a researcher, who needs access to BIMS, is to log into the system, search desired information (textual and images), and use them in his experiments. The BIMS architecture implementation provides web browser navigation, as a simple way to search information. For this reason, it would be necessary to develop a new module to make specific queries to BIMS, and extract all information. For example, a query could be to extract information (textual and images) about smoker patients, with systolic blood pressure between 12 and 15, into a local folder. The BIMS Query & Export Module would be responsible for executing the query and pack all generated information in a folder defined by the researcher.

**Standardized Reference Model**

The BIMS Reference Model is composed of a set of software objects to represent the model domain elements in a flexible way. A simple and proprietary architecture was developed, and even though it is a model adaptable to a wide range of possibilities, it is not based on standardized model. Currently, there are organizations, like CEN (European Committee for Standardization), which define specifications to develop and support global, platform-independent data standards (ISO 13606), that enable information system interoperability to improve medical research and related areas of health care. With the aim to provide interoperability features in the BIMS architecture, it would be necessary to modify, and adapt the BIMS Reference Model to a standard specification.

## Conclusions

This project was motivated by the need to address data acquisition problems commonly encountered in biomedical research projects. The main outcome has been BIMS (Biomedical Information Management System).

BIMS is a modular multilayer object–oriented software architecture, designed with the mission to provide a flexible infrastructure to capture, and manage large amounts of heterogeneous biomedical data sets, both textual as well as visual information. The flexibility is the most significant feature of BIMS architecture, because it provides the possibility to define, at run-time, a wide range of information needs commonly encountered in biomedical research studies, not being limited to any given clinical specialty. Therefore, BIMS can easily be adapted to store biomedical information about an extensive variety of projects.

The two-level modeling paradigm has been used to develop the BIMS architecture. Thus, it is possible to define, and model a wide range of biomedical concepts, without the need to modify underlying information system to fulfill new requirements. The use of a flexible system, in data acquisition and management, facilitates the job of clinicians and researchers, who usually encounters difficulties in data entry and data analysis [1]. Moreover, the BIMS architecture provides a set of mechanisms to improve data quality of the research project, with the aim to make its results more reliable.

In summary, BIMS meets two important goals in the management of biomedical information. On the one hand, it provides flexibility to manage a wide range of projects, facilitating the job of clinicians and researchers. On the other hand, BIMS provides a set of mechanisms to ensure a high data quality in the biomedical research projects.

## References

[1] Anderson N, Lee E, Brockenbrough J, Minie M, Fuller S, Brinkley J, Tarczy-Hornoch, P. Issues in biomedical research data management and analysis: Needs and barriers. J Am Med Inform Assn. 2007; 14(4): 478–488.

[2] Beale T. Archetypes: Constraint-based Domain Models for Future-proof Information Systems. Proceedings of the 11th OOPSLA Workshop Behavioral Semantics. 2002. p. 16-32.

[3] Bisbal J, Berry D. Archetype alignment. Two-level Driven Semantic Matching Approach to Interoperability in the Clinical Domain. Proceedings of the International Conference on Health Informatics. Porto: 2009. p. 216-221.

[4] Brazhnik O, F. Jones J. Anatomy of Data Integration. J Biomed Inform. 2007; 40(3): 252-269.

[5] Capdeferro D, Bisbal J, Frangi AF. Data Handling & Quality Challenges in a Multi-centre Biomedical Research Project. Medinfo 2010, submitted.

[6] Fetzer DT, West C. The HIPAA Privacy Rule and Protected Health Information Implications in Research Involving DICOM Image Databases. Acad Radiol. 2008; 15(3): 390-5.

[7] Garde S, Hovenga E, Buck J, Knaup P. Expressing clinical data sets with openEHR archetypes: a solid basis for ubiquitous computing. Int J Med Inform. 2007; 76(3): 334-41.

[8] Grimson J, Grimson W, Berry D, Stephens G, Felton E, Kalra D, Toussaint P, Weier OW. A CORBA-Based Integration of Distributed Electronic Healthcare Records Using the Synapses Approach. IEEE Trans. Inf. Tech. in Biomedicine. 1998; 2(3): 124-138.

[9] Martínez-Costa C, Menárguez-Tortosa M, Fernández-Breis JT, Maldonado JA. A model-driven approach for representing clinical archetypes for Semantic Web environments. J Biomed Inform. 2009; 42(1): 150-164.

[10] Sommerville I. Software Engineering. 7th edition. Addison Wesley; 2004.

**Address for correspondence**

E-mail address: oscarmoraperez@gmail.com